\documentclass[12pt]{article}

\renewcommand{\arraystretch}{1.5}
\renewcommand{\bar}{\overline}
\newcommand{\tr}{\mathop{\rm tr}}
\newcommand{\imag}{\mathop{\rm Im}}

\newcommand{\figsize}{\small}
\newcommand{\half}{{1\over 2}}
\newcommand{\stu}{SU(3)\times SU(2)\times U(1)}

\def\cA{{\cal A}}

\input prepictex
\input pictex
\input postpictex
\newdimen\tdim
\def\stpltsmbl{\setplotsymbol ({\small .})}

\def\tarrow{\arrow <7\tdim> [.3,.6]}
\newbox\sru
\setbox\sru=\hbox{\beginpicture
\setcoordinatesystem units <\tdim,\tdim>
\stpltsmbl
\setquadratic
\plot
  0.0   0.0
  4.8   1.5
  7.5   5.0
  7.3   8.5
  5.0  10.0
  2.7   8.5
  2.5   5.0
  5.2   1.5
 10.0   0.0
/
\endpicture}
\def\springru #1 #2 *#3 /{\multiput {\copy\sru}  at
#1 #2 *#3 10 0 /}
\newbox\srd
\setbox\srd=\hbox{\beginpicture
\setcoordinatesystem units <\tdim,\tdim>
\stpltsmbl
\setquadratic
\plot
  0.0   0.0
  4.8  -1.5
  7.5  -5.0
  7.3  -8.5
  5.0 -10.0
  2.7  -8.5
  2.5  -5.0
  5.2  -1.5
 10.0  -0.0
/
\endpicture}
\def\springrd #1 #2 *#3 /{\multiput {\copy\srd}  at
#1 #2 *#3 10 0 /}
\newbox\sdl
\setbox\sdl=\hbox{\beginpicture
\setcoordinatesystem units <\tdim,\tdim>
\stpltsmbl
\setquadratic
\plot
  0.0   0.0
 -1.5  -4.8
 -5.0  -7.5
 -8.5  -7.3
-10.0  -5.0
 -8.5  -2.7
 -5.0  -2.5
 -1.5  -5.2
 -0.0 -10.0
/
\endpicture}
\def\springdl #1 #2 *#3 /{\multiput {\copy\sdl}  at
#1 #2 *#3 0 -10 /}
\newbox\sdr
\setbox\sdr=\hbox{\beginpicture
\setcoordinatesystem units <\tdim,\tdim>
\stpltsmbl
\setquadratic
\plot
  0.0   0.0
  1.5  -4.8
  5.0  -7.5
  8.5  -7.3
 10.0  -5.0
  8.5  -2.7
  5.0  -2.5
  1.5  -5.2
  0.0 -10.0
/
\endpicture}
\def\springdr #1 #2 *#3 /{\multiput {\copy\sdr}  at
#1 #2 *#3 0 -10 /}

\newcommand{\beq}{\begin{equation}}
\newcommand{\eeq}{\end{equation}}

\begin{document}
\begin{titlepage}
\def\thepage {}        

\title{Soft Superweak CP Violation and the Strong CP Puzzle}

\author{
{\renewcommand{\arraystretch}{1}
\begin{tabular}{c}Howard Georgi\thanks{georgi@physics.harvard.edu}
~and Sheldon L. Glashow\thanks{glashow@physics.harvard.edu}
\\
Lyman Laboratory of Physics\\
Harvard University\\
Cambridge, MA 02138\end{tabular}}}

\date{7/98 revised 2/99}
\maketitle

\bigskip
\vspace{-3.3in}\begin{flushright}
{HUTP-98/A048}
\end{flushright}
\vspace{3.3in}

\begin{abstract} We discuss a class of models in which CP is violated
softly in a heavy sector adjoined to the standard model. Heavy-sector loops
produce the observed CP violation in kaon physics, yielding  a tiny
and probably undetectable value for $\epsilon^\prime$. All other
CP-violating parameters in the effective low-energy standard model,
including the area of the unitarity triangle and $\bar\theta$, are finite,
calculable and can be made very small. The leading contribution to
$\bar\theta$ comes from a four-loop graph.  These models offer a natural
realization of superweak CP violation and can resolve the strong CP puzzle. 
In one realization of this idea, CP is violated in the mass
matrix of heavy majorana neutrinos.

\pagestyle{empty}
\end{abstract}
\end{titlepage}

\setcounter{section}{0}

\section{Introduction}
\setcounter{equation}{0}

We examine the old idea of superweak CP violation~\cite{wolfenstein} from
the  modern perspective of naturalness. Superweak look-alike models are
easy to make: for example, by introducing  a second $SU(2)$-doublet
spinless field $\beta$, like the Higgs doublet  but without a vacuum
expectation value. When the couplings of the ordinary Higgs are real but those
of $\beta$ are complex, the exchange of a virtual $\beta$ is the sole
source of CP violation. If its couplings are so tiny that these
interactions are negligible except in the neutral $K$ mass matrix, we
obtain an effectively superweak model. 

However, models of this ilk are hideously unnatural.
There is no symmetry to keep the Yukawa
couplings of the Higgs boson real. Physically meaningful quantities like
the area of the unitarity triangle, or equivalently the Jarlskog $J$
parameter~\cite{jarlskog}, receive infinite contributions from quantum loop
effects. A free parameter of the theory --- the complex phase in the
Kobayashi-Maskawa (KM) matrix --- has been set to zero without any
justification.  We shall describe a class of superweak models that is free
of this  affliction, and as well, of a strong CP problem~\cite{strongcp}.
We show how to calculate the area of the unitarity triangle as a finite
radiative correction in these models. 

In our models, CP is violated softly in a heavy sector adjoined to an
otherwise standard and CP-conserving standard 
model.\footnote{Our models
can be adapted, with some additional architecture, to violate CP
spontaneously rather than softly. Some might find this more elegant, but
here we are striving for simplicity.}
Soft CP violation is an old idea\footnote{For other work on soft CP violation,
see \cite{bchao}, and references therein. This reference also describes models
which are quite similar to those we construct here. We will discuss the
differences below. But our focus is also somewhat different, in that we
concentrate on constucting models that are naturally superweak.}
We believe that the class of
models we describe in this paper is simpler than most models in the
literature, and more effective in
suppressing non-superweak effects. Our
low-energy sector consists solely
of the three fermion families, the $\stu$ gauge bosons, and one relic Higgs
boson. Because CP violation is soft, the dimension-4 interactions in the
model are naturally real. The only CP-violating phase appears in the mass
matrix of the heavy sector. In particular, the KM matrix is real at tree
level. CP-violating corrections to the dimension-4 couplings are generated
by quantum loops involving the heavy sector, but  they are finite and 
calculable functions of the renormalizable parameters of the model.
Observed CP violation in the neutral kaon system arises from a
dimension-6 interaction produced by a box diagram involving the heavy
particles. Two other important parameters relevant to potentially
observable CP-violating phenomena, the area of the unitarity triangle and
the strong CP violating parameter $\bar\theta$, are both tiny for a wide
choice of parameters: the former undetectable, the latter
innocuous. We calculate the leading contributions to both below.

Ours are classic superweak models~\cite{wolfenstein}\ where the only
observable effect of the new interactions is its contribution to the
CP-violating part of $K^0$-$\overline{K^0}$ mass mixing. This simple scheme
is disfavored  at the one-sigma level by current data on $B$
decay.\footnote{However, see the Note Added on page
\pageref{mele}.}
However, by choosing  the couplings of the heavy sector to the third family
larger than those to the first two families we can implement the ``3
scenario" of Barbieri, Hall, Stocci and Weiner (BHSW)~\cite{hall}.  In that
case, the superweak interactions can affect the neutral $B$-meson sector
so as to fit the data better. We do not discuss this possibility here.

\section{A Simple Model}

We append two new and heavy particle species  to the 
the standard model: two multiplets of one sort, $\xi_\alpha$,
where $\alpha=1$ or 2,
and one of another, $\chi$. 
We begin by specifying only two properties of these particles:
\begin{enumerate}
\item CP is violated softly in the $\xi$ mass matrix, {\it i.e.,} by
dimension-2 operators if $\xi$ is a boson, or by dimension-3
operators if it is a fermion.
\item There are renormalizable Yukawa couplings by which
$\xi_\alpha$ and $\chi$ couple in pairs to the
light left-handed quark doublets $\psi_L$ as in
 figure~\ref{fig-1}, where:
\begin{equation}
\psi_L=\pmatrix{
V^\dagger\,U_L\cr
D_L\cr}\,.
\label{lhd}
\end{equation} 
$V$, the tree-level KM matrix, is real because of our
assumption of soft CP violation.
It follows that $\xi$ is a spinless boson and $\chi$ a spin ${1\over
2}$ fermion, or {\it vice versa.} We will argue later that we obtain the
maximum suppression of $\bar\theta$ if the $\xi_\alpha$ are colorless
fermions, but for now, we will leave this unspecified.
\end{enumerate}
We denote the
Feynman amplitude for the coupling as $f\,o_{i\alpha}$, where $i=d,s,b$ is a
flavor index. An overall constant $f$ sets the scale of the
Yukawa couplings, and we will assume the components of $o_{i\alpha}$ to be
of order unity.\footnote{In the 3 scenario of \cite{hall}, the $i=b$
components
$o_{i\alpha}$ would be larger that the others.}
In this basis, the assumption of soft CP breaking requires 
$o_{i\alpha}$ to be real. 

{\figsize\begin{figure}[htb]
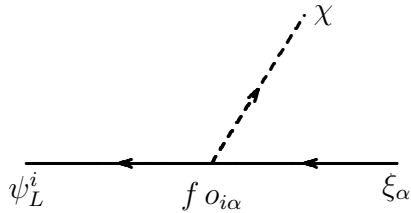

$$\beginpicture
\setcoordinatesystem units <\tdim,\tdim>
\stpltsmbl
\tarrow from 100 0 to 50 0
\tarrow from 0 0 to -50 0
\plot 100 0 -100 0 /
\put {$\psi^i_L$} [t] at -100 -5 
\put {$\xi_\alpha$} [t] at 100 -5
\tarrow from 22.5 36 to 25 40 
\setdashes <3pt>
\tarrow from 0 0 to 25 40
\plot 0 0 50 80 /
\put {$\chi$} [l] at 55 80
\put {$f\,o_{i\alpha}$} [t] at 0 -10  
\endpicture$$
\caption{\figsize\label{fig-1} The coupling of the left-handed quark doublet
to the heavy sector ($i$ is a flavor index). The $o_{i\alpha}$ are
real.}\end{figure}}

{\figsize\begin{figure}[htb]
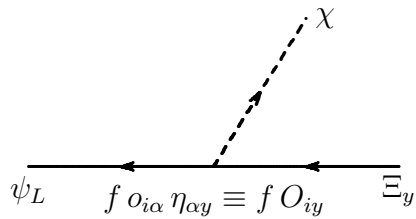

$$\beginpicture
\setcoordinatesystem units <\tdim,\tdim>
\stpltsmbl
\tarrow from 100 0 to 50 0
\tarrow from 0 0 to -50 0
\plot 100 0 -100 0 /
\put {$\psi_L$} [t] at -100 -5 
\put {$\Xi_y$} [t] at 100 -5
\tarrow from 22.5 36 to 25 40 
\setdashes <3pt>
\tarrow from 0 0 to 25 40 
\plot 0 0 50 80 /
\put {$\chi$} [l] at 55 80
\put {$\displaystyle f\,o_{i\alpha}\,\eta_{\alpha y}
\equiv f\,O_{iy}$} [t] at 0 -10
\endpicture$$
\caption{\figsize\label{fig-2}The coupling of the left-handed quarks to the
heavy
sector in the mass eigenstate basis for the $\Xi$'s. The $O_{ix}$ are
complex.}\end{figure}}

Soft CP violation in the $\xi$ mass matrix leads to mass
eigenstates $\Xi_y$ that are
related to $\xi_\alpha$ by a (complex) unitary
transformation, 
\begin{equation}
\xi_\alpha= \eta_{\alpha y}\,\Xi_y\,.
\label{fd1}
\end{equation}
It is convenient to express the new interaction in terms of these mass
eigenstates. This is illustrated in figure~\ref{fig-2}, 
in which we have defined 
\begin{equation}
O_{iy}\equiv \sum_\alpha \,o_{i\alpha}\,\eta_{\alpha y}\,.
\label{fd2}
\end{equation}
If the $\xi$s are scalars, the field redefinition (\ref{fd1}) can generate
phases in their self-couplings. These do not affect
low-energy physics directly, and their virtual effects inside heavy sector
loops are generally small compared to the effects due to the complex
parameters $O_{iy}$. Thus, we analyze
the CP violation  arising exclusively from these 
parameters, which
can be done without considering the details of the heavy sector
physics. 

Arrows placed on the heavy sector lines in figure~\ref{fig-1} and
figure~\ref{fig-2} represent the flow of a quantum number, ``heaviness'',
associated with the heavy sector fields. The interactions of
figure~\ref{fig-1} and figure~\ref{fig-2} conserve this quantum number. It
could be that there are weaker interactions that violate heaviness, but we
will not discuss these.  If either $\Xi_x$ or $\chi$ are neutral Majorana
particles,  heaviness would only be conserved modulo 2. In these special
cases (which we discuss separately) there are additional contributions to
CP-violating effects. The arrows in figure~\ref{fig-1} are important for
another reason. In our analysis of the CP violation in these models, we
will sometimes treat the $x$ index on $\Xi$ as an additional flavor. The
structure of figure~\ref{fig-2} is such that this generalized flavor flows
along the solid line with the CP-violating coupling $f\,O_{ix}$  regarded
as a matrix in flavor space.

{\figsize\begin{figure}[htb]
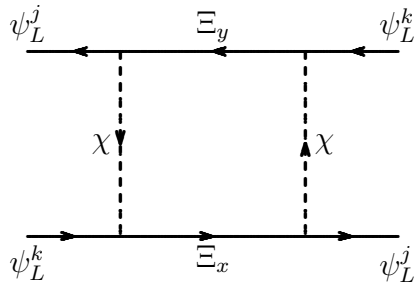

$$\beginpicture
\setcoordinatesystem units <\tdim,\tdim>
\stpltsmbl
\plot 100 50 -100 50 / 
\plot 100 -50 -100 -50 /
\tarrow from 80 50 to 75 50 
\tarrow from 5 50 to 0 50 
\tarrow from -70 50 to -75 50
\tarrow from 70 -50 to 75 -50 
\tarrow from -5 -50 to -0 -50 
\tarrow from -80 -50 to -75 -50
\tarrow from -50 5 to -50 0 
\tarrow from 50 -5 to 50 0 
\setdashes <3pt>
\plot 50 50 50 -50 /
\plot -50 50 -50 -50 /
\put {$\psi_L^k$} [t] at -100 -55 
\put {$\Xi_x$} [t] at 0 -55
\put {$\Xi_y$} [b] at 0 55
\put {$\psi_L^j$} [t] at 100 -55 
\put {$\psi_L^k$} [b] at 100 55 
\put {$\psi_L^j$} [b] at -100 55
\put {$\chi$} [r] at -55 0
\put {$\chi$} [l] at 55 0
\endpicture$$
\caption{\figsize\label{fig-3}This box graph produces the superweak
interaction.}\end{figure}}
The superweak interaction arises from the box graph shown in
figure~\ref{fig-3}. The coefficient of the effective 4-fermion
coupling produced by figure~\ref{fig-3} at the scale $M_\chi$ is
of order:
\begin{equation}
{\alpha_f^2\,I_{xy}\over M_\chi^2}\,
O_{jx}\,O_{kx}^{*}
\,O_{jy}\,O_{ky}^{*}
+ x \leftrightarrow y
\label{sw1}
\end{equation}
where $(j,k)=(d,s)$, $(d,b)$, or $(b,s)$ and where $\alpha_f={f^2/4\pi}$.
$I_{xy}$ is an integral  depending on $m_{\Xi_x}^2/M_\chi^2$ and
$m_{\Xi_y}^2/M_\chi^2$. If the dimensional parameters are the same order of
magnitude, these mass ratios and $I_{xy}$ are of order unity. The details
of the 4-fermion operator produced depend on the specific properties of
$\xi_\alpha$ and $\chi$ but all lead to effective flavor-changing  
4-fermion
interactions.  The operator from figure~\ref{fig-3} is
renormalized by QCD effects at lower energies, but this effect is less
than a factor of two in all circumstances we consider. It does not affect
our order of magnitude estimates and will be ignored. Our models, as we
shall demonstrate, are natural realizations of the superweak model. Thus the
interaction corresponding to  figure~\ref{fig-3}, with 
$j=d$ and  $k=s$ (two incoming $s$
quarks and two exiting $d$  quarks), provides essentially all
the observed CP violation in the neutral kaon system.

\section{The area of the unitarity triangle}

The CP-violating corrections to the renormalizable
interactions of the low-energy standard model must be proven to be small
if figure~\ref{fig-3} is to be responsible for all observable CP
violation in our model.
In particular, radiative corrections will induce
finite complex phases in the (initially real)
KM matrix. Some of these phases can be removed by field redefinitions and do
not represent real CP violation. But the area of the unitarity triangle,
\begin{equation}
\cA
\equiv\half\biggl|\imag\Bigl(V_{ub}\,V_{ud}^*\,V_{cb}^*\,V_{cd}\Bigr)\Biggr|
\,,
\label{ut1}
\end{equation}
is an invariant measure of the KM CP violation. In the standard model,
and in the standard parameterization used in the particle data booklet, it is
\begin{equation}
\cA=\half\Bigl|\sin\delta_{13}\,s_{12}\,s_{13}\,s_{23}\,c_{12}\,c_{13}^2\,
c_{23}\Bigr|\,.
\label{ut3}
\end{equation}

{\figsize\begin{figure}[htb]
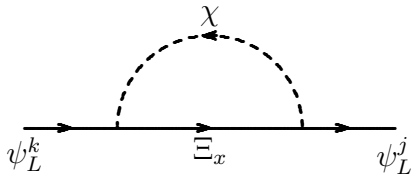

$$\beginpicture
\setcoordinatesystem units <\tdim,\tdim>
\stpltsmbl
\plot 100 -50 -100 -50 /
\tarrow from 70 -50 to 75 -50 
\tarrow from -5 -50 to -0 -50 
\tarrow from -80 -50 to -75 -50
\tarrow from 3.5 0 to -3.5 0 
\setdashes <3pt>
\put {$\psi_L^k$} [t] at -100 -55 
\put {$\Xi_x$} [t] at 0 -55
\put {$\psi_L^j$} [t] at 100 -55 
\put {$\chi$} [b] at 0 5
\circulararc 180 degrees from 50 -50 center at 0 -50
\endpicture$$
\caption{\figsize\label{fig-4}This fermion self-energy graph introduces phases
into the KM matrix.}\end{figure}}

The leading contribution to $\cA$ comes from the self-energy diagram in
figure~\ref{fig-4}, whose
CP-conserving part produces an irrelevant infinite
renormalization of the real parameters of the theory. Its
CP-violating part is finite, calculable, and produces real physical effects. 
This contribution has the form
\begin{equation}
\bar{\psi_L^j}\,\,i\!\not\!\! D\,\,iC_{jk}\,\psi_L^k
\label{ut4}
\end{equation}
where $C$ is a real, antisymmetric matrix in flavor space. Up to factors of
order one (depending on the $\Xi-\chi$ mass ratios), it is
\begin{equation}
C_{jk}\approx
{\alpha_f\over4\pi}\,
\imag\left(O_{j1}\,O_{k1}^*\right)
\approx {\alpha_f\over4\pi}\,
\left(o_{j1}o_{k2}-o_{j2}o_{k1}\right)\,.
\label{ut5}
\end{equation}
We must redefine the $\psi_L$ field to 
restore the canonical form of its kinetic energy. 
Hereafter we assume that $\alpha_f$ (and hence $C$) is small.
To lowest order in $\alpha_f$, the required field
redefinition is:
\begin{equation}
\psi_L\rightarrow \psi'_L=(I+iC/2)\,\psi_L\,.
\label{ut6}
\end{equation}  
The complex field redefinition, (\ref{ut6}), also introduces phases into the
quark mass matrices, and thus into the KM matrix.
The Yukawa couplings that give rise to the tree level quark masses have the
form
\begin{equation}
{\sqrt2\over v}\,\bar{D_R}\,\phi^\dagger\,M_D\,\psi_L
+{\sqrt2\over v}\,\bar{U_R}\,{\tilde\phi}^\dagger\,M_U\,V\,\psi_L
+{\rm h.c.}
\label{ut7}
\end{equation}
In terms of the redefined fields, these Yukawa couplings become
\begin{equation}
{\sqrt2\over v}\,\bar{D_R}\,\phi^\dagger\,M_D\,(I-iC/2)\,\psi'_L
+{\sqrt2\over v}\,\bar{U_R}\,\tilde\phi^\dagger\,M_U\,V\,(I-iC/2)\,\psi'_L
+{\rm h.c.}
\label{ut72}
\end{equation}
and lead to the radiatively-corrected mass terms:
\begin{equation}
\bar{D_R}\,M_D\,(I-iC/2)\,D'_L
+\bar{U_R}\,M_U\,V\,(I-iC/2)\,V^\dagger\,U'_L
+{\rm h.c.}
\label{ut73}
\end{equation}
where $M_D$ and $M_U$ are diagonal matrices.
To lowest order, this field redefinition does not affect the quark mass
eigenvalues, but it does change the KM matrix.

To estimate $\cA$ we must determine the correction to $V$, the
tree-level KM matrix.
As a first step, we diagonalize the mass-squared matrices of the
left-handed quarks, which from (\ref{ut73}) are:
\begin{equation}
(I-iC/2)\,M^2_D\,(I-iC/2)\quad\ {\rm and}\quad\ (I-iC^\prime/2)\,
          M^2_U\,(I-iC^\prime/2)\,, 
\label{ut8}
\end{equation}
where
\begin{equation}
C^\prime=V\,C\,V^\dagger\,.
\label{ut102}
\end{equation}
These mass-squared matrices may be diagonalized by unitary
transformations
Again to order $\alpha_f$, the appropriate transformations are:
\begin{equation}
D_L^{\prime \prime}=(I+iF)\,D'_L
\quad\ {\rm and}\quad\ 
U_L^{\prime\prime}=(I+iG)\,U'_L
\label{ut9}
\end{equation}
where $F^\dagger=F$ and $G^\dagger=G$ are real symmetric matrices. To lowest
order in $C$ they are:
\begin{equation}
F_{jk}=-\half\,C_{jk}\,{m_j^2+m_k^2\over m_j^2-m_k^2}
\quad\ {\rm and}\quad\ 
G_{jk}=-\half\,C'_{jk}\,{m_j^2+m_k^2\over m_j^2-m_k^2}\,,
\label{ut10}
\end{equation}
for $j\ne k$ and $F_{jk}=G_{jk}=0$ otherwise.
The KM
matrix, corrected  for the effect of figure~\ref{fig-4}, is:
\begin{equation}
(1-iG)\,V\,(1+iF)\,.
\label{ut11}
\end{equation}

We proceed to calculate the area of the unitarity triangle. If $\cA$
were zero, all phases in the KM matrix could be removed by field
redefinitions
and there would be no CP violation from $W$ exchange. 
If it were very small compared to its standard-model value, (\ref{ut3}),
{\it i.e.,} if
\begin{equation}
\cA\ll |s_{12}\,s_{13}\,s_{23}|\,,
\label{ut12}
\end{equation}
the remnant phase in the KM matrix would also be very small and
CP violation from $W$ exchange could  be neglected.
Explicit calculation to lowest order in $C$
gives
{\renewcommand{\arraystretch}{2.5}
\begin{equation}
\begin{array}{c}
\displaystyle
\cA=\half\,
\left|
-s_{12}\,s_{13}^2\,C_{12}+s_{13}^2\,s_{23}\,C_{23}
+\left({m_d^2\over m_s^2}-{m_u^2\over m_c^2}\right)\,s_{13}\,s_{23}\,C_{12}
\right.\\
\displaystyle
\left.
-\left({m_d^2\over m_b^2}-{m_u^2\over m_t^2}\right)\,s_{12}\,s_{23}\,C_{13}
+\left({m_s^2\over m_b^2}-{m_c^2\over m_t^2}\right)\,s_{12}\,s_{13}\,C_{23}
+\cdots
\right|
\end{array}
\label{ut13}
\end{equation}}%
\noindent
where $\cdots$ indicate terms that
are suppressed by additional factors of small quark mass
ratios.
The terms in (\ref{ut13}) are proportional to elements of the matrix $C$.
Each coefficient is suppressed relative to the right-hand side of
(\ref{ut12}), either by small quark mass ratios or because $s_{13}$ is 
the smallest of the KM angles. Thus the area of the unitarity model is tiny
compared to its standard-model value unless $\alpha_f\ge 4\pi$.
We have succeeded
in producing a model in which all CP violation from $W$ exchange is naturally
negligible. 

\section{Strong CP}

The basic idea of soft CP violation as a solution to the strong CP problem is
that CP invariance in the absence of soft breaking requires that the QCD
$\theta$ parameter be zero. Then all CP violating corrections to $\bar\theta$
due to the soft breaking are finite and calculable in terms of the soft
breaking parameters. One can check whether they are small enough avoid
phenomenological problems. In some realizations of our models, we will see
that significant strong CP violation is induced. Others will survive this
hurdle.

We first consider contributions to $\bar\theta$ in tree approximation. These
can arise only if there are colored fermions that have CP violating mass
matrices in tree approximation. But if the $\Xi_x$ are colored fermions,
it would be unnatural for the phase of the determinant
of theie mass matrix should vanish. To avoid a strong CP puzzle at the tree
level, we must assume that the $\Xi_x$ are not colored fermions.\footnote{This
was emphasized to us by Darwin Chang. See \cite{bchao}.}
We hereafter suppose that the $\Xi_x$ are either scalars or colorless, and
consider loop diagrams that could contribute to $\bar\theta$.

The field redefinitions produced by CP violating self-energy diagrams
like that in figure~\ref{fig-4} do not induce a non-zero value of $\bar\theta$
to
any order in $\alpha_f$ because they are necessarily hermitian in flavor space
(they are associated with hermitian counter-terms in the Lagrangian).
Therefore the determinant of the transformed mass matrix
is real. This result applies to the field redefinitions produced by any
self-energy diagram, however complex.

CP-violating loop corrections to the Yukawa couplings can generate non-zero
$\bar\theta$.  These corrections begin at the two-loop level. However, we
shall
see that if we add one additional restriction, our models require a rather
complicated flavor structure to produce a phase in the determinant of the
quark mass matrix, so that the leading contributions to $\bar\theta$ are
very small.  We shall identify the required flavor structure, then find and
estimate those Feynman graphs giving the largest contribution. 

We assume that the heavy-sector masses are large compared to the TeV scale
of electroweak symmetry breaking. Thus quark masses  appearing in the
denominators of Feynman integrals can be ignored, but explicit quark mass
dependence  arises from the Yukawa couplings (\ref{ut7}) of the Higgs
doublet $\phi$ which, in our models, are real at tree level.
Suppose heavy sector loops were to produce a complex
CP-violating contribution to the
Yukawa couplings of the form:
\begin{equation}
{\sqrt2\over v}\,\bar{D_R}\,\phi^\dagger\,\Delta M_D\,\psi_L
+{\sqrt2\over v}\,\bar{U_R}\,{\tilde\phi}^\dagger\,\Delta M_U\,V\,\psi_L
+{\rm h.c.}
\label{hc2}
\end{equation}
Its lowest-order contribution to the phase of the determinant of the
quark masses is
\begin{equation}
\Delta\bar\theta\approx\imag\biggl[\tr\biggl(
\Delta M_D\,M_D^{-1}
+\Delta M_U\,M_U^{-1}\,. 
\biggr)\biggr]
\label{hc3}
\end{equation}
Because some quark masses are small, the inverse quark mass matrices in
(\ref{hc3}) threaten to produce large phases. However, no such large
effects occur in our model. The heavy-sector particles couple only to the
left-handed quarks. The only couplings of the right-handed quarks are the
Yukawa couplings of (\ref{ut7}), which themselves are proportional to the
quark mass matrices. Thus the inverse mass matrices in (\ref{hc3}) are
always canceled. Had we instead coupled the heavy sector to the
right-handed quarks,  this cancellation would not occur and there would be
a potentially larger contributions to $\bar\theta$, in only two loops. We
discuss this
contribution in the Appendix.

Another potential two-loop contribution to $\bar\theta$
imposes an important constraint on our class of models. If the $\Xi_x$ are
scalars, there are renormalizable couplings of the $\Xi_x$ to the Higgs
doublet, and we can draw the diagram of figure~\ref{fig-higgs}~\cite{bchao}.
This contribution is proportional to the unknown coupling
constant for the coupling of two $\Xi$s to $\phi^\dagger\phi$. We could
assume that this coupling is small and ignore it, but it seems more elegant to
eliminate the diagram entirely by assuming that the $\Xi_x$ are fermions,
which we do in the remainder of this note.

{\figsize\begin{figure}[htb]
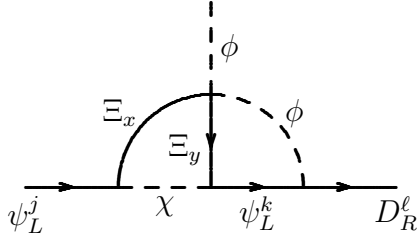

$$\beginpicture
\setcoordinatesystem units <\tdim,\tdim>
\stpltsmbl
\plot 100 -50 0 -50 /
\plot -50 -50 -100 -50 /
\tarrow from 70 -50 to 75 -50 
\tarrow from 0 -24 to 0 -28 
\tarrow from 24 -50 to 28 -50 
\tarrow from -80 -50 to -75 -50
\setdashes <5pt>
\plot 0 0 0 50 /
\plot 0 -50 -50 -50 /
\put {$\psi_L^j$} [t] at -100 -55 
\put {$\psi_L^k$} [t] at 25 -55
\put {$\chi$} [t] at -25 -55
\put {$D_R^\ell$} [t] at 100 -55 
\put {$\phi$} [l] at 5 25
\put {$\phi$} [l] at 40 -10
\put {$\Xi_x$} [r] at -40 -10
\put {$\Xi_y$} [r] at -5 -30
\circulararc 90 degrees from 50 -50 center at 0 -50
\setsolid
\circulararc -90 degrees from -50 -50 center at 0 -50
\plot 0 0 0 -50 /
\linethickness=0pt
\putrule from 0 60 to 0 -70 
\endpicture$$
\caption{\figsize\label{fig-higgs}A contribution to the Yukawa
coupling that contributes to $\bar\theta$ if the $\Xi_x$ are
scalars.}\end{figure}}

To simplify our study of other consequences of (\ref{hc3}), we make use of
the essential notion mentioned earlier, of $x$ as a generalized flavor
index allowing us to trace flavor through each Feynman diagram. For example,
the graph in figure~\ref{fig-4} is proportional to the
hermitian  flavor matrix $O_x\,O_x^\dagger$.  

We first show that diagrams
involving heavy sector loops, but no Higgs loops, do not contribute to
$\bar\theta$. These diagrams do not involve quark mass matrices
because the inverse mass matrix in (\ref{hc3}) cancels the quark mass
matrix from the Yukawa coupling. Because $\Delta \bar{\theta}$  is a trace
in flavor space, the quark flavor indices are always summed over. Thus the
trace is a product of objects of the following form:
\begin{equation}
K_{xy}=\sum_j\,O_{jx}^*\,O_{jy}
\label{tb4}
\end{equation}
where $K_{11}$ and $K_{22}$ are real and  $K_{12}=K_{21}^{\dagger}$.
Because $x$ flavor is conserved,
$x$ indices  appear at each end of every $x$ line:
once as the first index of a $K_{xy}$ factor and once as the second.
Thus the trace involves equal numbers of
$K_{12}$ and $K_{21}$ factors and is necessarily 
real. Had we introduced
three $\Xi$s rather than two, figure~\ref{fig-7} would give a
non-zero contribution $\sim \alpha_f^3 \alpha_s/(4\pi)^4$
proportional to the possibly-complex product $K_{12}\,K_{23}\,K_{31}$.
The gluon loop in figure~\ref{fig-7} is necessary to
produce a one-particle-irreducible contribution to the Yukawa
coupling.

{\figsize\begin{figure}[htb]
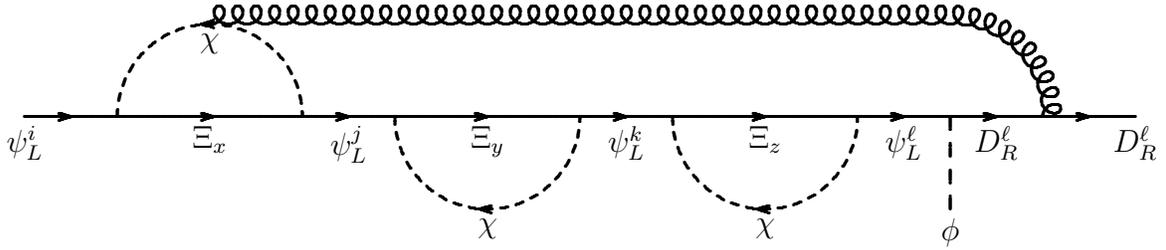

$$\beginpicture
\setcoordinatesystem units <\tdim,\tdim>
\stpltsmbl
\plot 200 -50 -400 -50 /
\tarrow from 70 -50 to 75 -50 
\tarrow from -5 -50 to -0 -50 
\tarrow from -80 -50 to -75 -50
\tarrow from 120 -50 to 125 -50 
\tarrow from 170 -50 to 175 -50 
\tarrow from -155 -50 to -150 -50
\tarrow from -230 -50 to -225 -50
\tarrow from -303 -50 to -298 -50
\tarrow from -380 -50 to -375 -50
\tarrow from 1.5 -100 to -4.5 -100 
\tarrow from -148.5 -100 to -154.5 -100 
\tarrow from -298.5 0 to -304.5 0 
\setquadratic
\plot
150.0 -50.0
151.3 -46.1
154.6 -43.5
158.1 -43.3
159.8 -45.3
158.5 -47.6
155.0 -47.9
151.3 -45.8
149.4 -42.2
150.1 -38.1
152.9 -35.0
156.4 -34.2
158.3 -36.0
157.4 -38.4
154.0 -39.3
150.0 -37.9
147.6 -34.5
147.6 -30.4
149.9 -26.9
153.2 -25.6
155.4 -27.0
154.9 -29.6
151.6 -31.0
147.5 -30.2
144.6 -27.3
143.9 -23.2
145.7 -19.4
148.7 -17.6
151.2 -18.7
151.0 -21.3
148.0 -23.2
143.8 -23.0
140.5 -20.6
139.2 -16.7
140.3 -12.6
143.1 -10.4
145.6 -11.0
145.9 -13.7
143.2 -16.0
139.1 -16.5
135.4 -14.6
133.5 -10.9
134.0  -6.8
136.3  -4.1
139.0  -4.4
139.6  -6.9
137.4  -9.7
133.3 -10.8
129.4  -9.5
127.0  -6.2
126.8  -2.0
128.7   1.0
131.3   1.2
132.4  -1.3
130.6  -4.3
126.8  -6.1
122.7  -5.4
119.8  -2.5
119.0   1.6
120.4   4.9
123.0   5.4
124.4   3.2
123.1  -0.1
119.6  -2.4
115.5  -2.4
112.1   0.0
110.7   4.0
111.6   7.4
114.0   8.3
115.8   6.4
115.0   2.9
111.9   0.1
107.8  -0.6
104.2   1.3
102.1   5.0
102.4   8.5
104.7   9.8
106.7   8.1
106.5   4.6
103.9   1.3
100.0   0.0
/
\springru 60 0 *3 /
\springru -300 0 *35 /
\setlinear
\setdashes <3pt>
\circulararc 180 degrees from -50 -50 center at 0 -50
\circulararc 180 degrees from -200 -50 center at -150 -50
\circulararc -180 degrees from -350 -50 center at -300 -50
\put {$\psi_L^i$} [t] at -400 -55 
\put {$\psi_L^j$} [t] at -225 -55 
\put {$\Xi_x$} [t] at -300 -55
\put {$\Xi_y$} [t] at -150 -55
\put {$\psi_L^k$} [t] at -75 -55 
\put {$\Xi_z$} [t] at 0 -55
\put {$\psi_L^\ell$} [t] at 75 -55 
\put {$D_R^\ell$} [t] at 125 -55 
\put {$D_R^\ell$} [t] at 200 -55 
\put {$\chi$} [t] at 0 -105
\put {$\chi$} [t] at -150 -105
\put {$\chi$} [t] at -300 -5
\put {~} at 0 0
\setdashes <5pt>
\plot 100 -100 100 -50 /
\put {$\phi$} [t] at 100 -105
\endpicture$$
\caption{\figsize\label{fig-7}A contribution to the Yukawa
coupling that could produce a contribution to $\bar\theta$ were there more
than two $\Xi$s.}\end{figure}}

{\figsize\begin{figure}[htb]
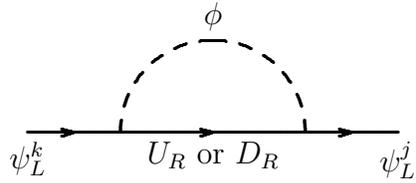

$$\beginpicture
\setcoordinatesystem units <\tdim,\tdim>
\stpltsmbl
\plot 100 -50 -100 -50 /
\tarrow from 70 -50 to 75 -50 
\tarrow from -5 -50 to -0 -50 
\tarrow from -80 -50 to -75 -50
\setdashes <5pt>
\put {$\psi_L^k$} [t] at -100 -55 
\put {$U_R$ or $D_R$} [t] at 0 -55
\put {$\psi_L^j$} [t] at 100 -55 
\put {$\phi$} [b] at 0 5
\circulararc 90 degrees from 50 -50 center at 0 -50
\circulararc -90 degrees from -50 -50 center at 0 -50
\endpicture$$
\caption{\figsize\label{fig-6}A Higgs loop that introduces dependence on the
quark mass matrices.}\end{figure}}

To find a contribution to $\bar\theta$, we must consider diagrams involving
two different non-trivial
flavor structures. This is possible in the presence of Higgs loops
that introduce quark mass-matrix dependence. However, it is not enough to have
one Higgs loop and one heavy sector loop. 
A Higgs
loop like that shown in figure~\ref{fig-6}
produces a contribution with flavor structure: 
\begin{equation}
B_{jk}\equiv\biggl[V^T\,M_U^2\,V+M_D^2\biggr]_{jk}\,,
\label{higgsloop}
\end{equation}
while a $\Xi$ loop like that shown in
figure~\ref{fig-4} has flavor structure: 
\begin{equation}
W^x_{jk}\equiv O_{jx}\,O^*_{kx}
\label{xiloop}
\end{equation}
for $x=1$ or 2.
Both are hermitian matrices so that the trace of their product 
is real. To have a complex trace, the diagram must involve the three
independent hermitian matrices $B$, $W^1$ and $W^2$. This structure arises
from graphs with two $\Xi$ loops, one Higgs loop, and a gluon loop to make
them one-particle irreducible, like that shown in figure~\ref{fig-7a}.

{\figsize\begin{figure}[htb]
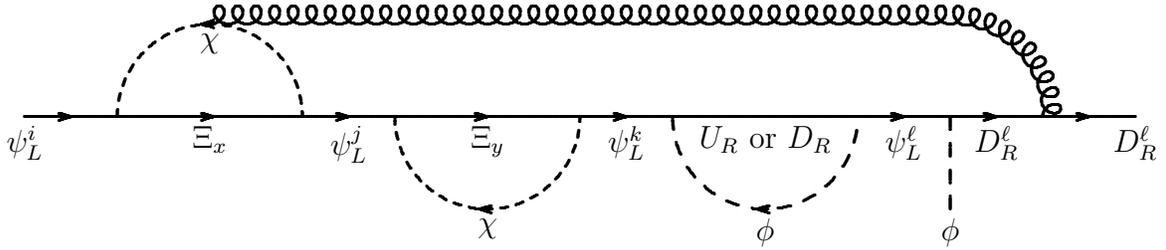

$$\beginpicture
\setcoordinatesystem units <\tdim,\tdim>
\stpltsmbl
\plot 200 -50 -400 -50 /
\tarrow from 70 -50 to 75 -50 
\tarrow from -80 -50 to -75 -50
\tarrow from 120 -50 to 125 -50 
\tarrow from 170 -50 to 175 -50 
\tarrow from -155 -50 to -150 -50
\tarrow from -230 -50 to -225 -50
\tarrow from -303 -50 to -298 -50
\tarrow from -380 -50 to -375 -50
\tarrow from 1.5 -100 to -4.5 -100 
\tarrow from -148.5 -100 to -154.5 -100 
\tarrow from -298.5 0 to -304.5 0 
\setquadratic
\plot
150.0 -50.0
151.3 -46.1
154.6 -43.5
158.1 -43.3
159.8 -45.3
158.5 -47.6
155.0 -47.9
151.3 -45.8
149.4 -42.2
150.1 -38.1
152.9 -35.0
156.4 -34.2
158.3 -36.0
157.4 -38.4
154.0 -39.3
150.0 -37.9
147.6 -34.5
147.6 -30.4
149.9 -26.9
153.2 -25.6
155.4 -27.0
154.9 -29.6
151.6 -31.0
147.5 -30.2
144.6 -27.3
143.9 -23.2
145.7 -19.4
148.7 -17.6
151.2 -18.7
151.0 -21.3
148.0 -23.2
143.8 -23.0
140.5 -20.6
139.2 -16.7
140.3 -12.6
143.1 -10.4
145.6 -11.0
145.9 -13.7
143.2 -16.0
139.1 -16.5
135.4 -14.6
133.5 -10.9
134.0  -6.8
136.3  -4.1
139.0  -4.4
139.6  -6.9
137.4  -9.7
133.3 -10.8
129.4  -9.5
127.0  -6.2
126.8  -2.0
128.7   1.0
131.3   1.2
132.4  -1.3
130.6  -4.3
126.8  -6.1
122.7  -5.4
119.8  -2.5
119.0   1.6
120.4   4.9
123.0   5.4
124.4   3.2
123.1  -0.1
119.6  -2.4
115.5  -2.4
112.1   0.0
110.7   4.0
111.6   7.4
114.0   8.3
115.8   6.4
115.0   2.9
111.9   0.1
107.8  -0.6
104.2   1.3
102.1   5.0
102.4   8.5
104.7   9.8
106.7   8.1
106.5   4.6
103.9   1.3
100.0   0.0
/
\springru 60 0 *3 /
\springru -300 0 *35 /
\setlinear
\setdashes <3pt>
\circulararc 180 degrees from -200 -50 center at -150 -50
\circulararc -180 degrees from -350 -50 center at -300 -50
\put {$\psi_L^i$} [t] at -400 -55 
\put {$\psi_L^j$} [t] at -225 -55 
\put {$\Xi_x$} [t] at -300 -55
\put {$\Xi_y$} [t] at -150 -55
\put {$\psi_L^k$} [t] at -75 -55 
\put {$U_R$ or $D_R$} [t] at 0 -55
\put {$\psi_L^\ell$} [t] at 75 -55 
\put {$D_R^\ell$} [t] at 125 -55 
\put {$D_R^\ell$} [t] at 200 -55 
\put {$\phi$} [t] at 0 -105
\put {$\chi$} [t] at -150 -105
\put {$\chi$} [t] at -300 -5
\put {~} at 0 0
\setdashes <5pt>
\plot 100 -100 100 -50 /
\circulararc 180 degrees from -50 -50 center at 0 -50
\put {$\phi$} [t] at 100 -105
\endpicture$$
\caption{\figsize\label{fig-7a}A contribution to the Yukawa
coupling that produces a non-zero $\bar\theta$.}\end{figure}}
The largest contribution to $\bar\theta$ from such Feynman diagrams
is of order
\begin{equation}
{\alpha_s\over4\pi}\,\left({\alpha_f\over4\pi}\right)^2\,
{V^T\,M_U^2\,V+M_D^2\over16\pi^2 v^2}
\approx{\alpha_s\over4\pi}\,\left({\alpha_f\over4\pi}\right)^2\,
{m_t^2\over16\pi^2 v^2}\,,
\label{thetabar1}
\end{equation}
where the simple form is sufficient because the $t$ quark is so much
heavier than the others. 

\section{Numbers}

The ratio  $\alpha_f/M_\chi$ is determined by the requirement that the graph
in
figure~\ref{fig-3} reproduces the observed CP violation observed in 
neutral kaons. We set equal the
contribution from the four-fermion operator $(\bar{d_L}\gamma^\mu
s_L)\,(\bar{d_L}\gamma_\mu s_L)$ with coefficient (\ref{sw1}) and the
product $\epsilon \Delta m_K$, where
$\epsilon\approx 2\times10^{-3}$ and
$\Delta m_K\approx3.52\times10^{-15}$~GeV. To evaluate the contribution from
the four-fermion operator, we use the vacuum insertion approximation, not
because we believe it, but because it is simple and likely to yield 
the correct order of
magnitude. We find:
\begin{equation}
\epsilon\,\Delta m_K\approx
{1\over2m_K}\,{8\over3}f_k^2m_K^2\,{\alpha_f^2\over M_\chi^2}
\label{bound1}
\end{equation}
or
\begin{equation}
{\alpha_f\over M_\chi}\approx 2\times10^{-8}\,{\rm GeV}^{-1}\,.
\label{bound2}
\end{equation}

To resolve the strong CP problem, it is necessary that (\ref{thetabar1}),
the dominant contribution to $\bar\theta$ in our models, 
is less than the bound of $\bar{\theta}<3\times10^{-10}$ following
from searches for
the neutron electric dipole moment~\cite{strongcp}. This yields the
constraint:
\begin{equation}
\alpha_f<0.044\,.
\label{thetabar2}
\end{equation}
Combining (\ref{bound2}) and (\ref{thetabar2}), we find:
\begin{equation}
M_\chi<2\times10^6\,{\rm GeV}\,.
\label{thetabar3}
\end{equation}
This upper bound on $M_\chi$ 
is comfortably above the electroweak breaking scale, so our
assumption that the heavy sector is far removed from the low-energy sector
is justified. 
Evidently, there is plenty of room for a natural superweak model that
solves the strong CP problem. 

\section{Majorana Fermions}

We consider the interesting special case where either $\chi$ or the $\Xi_x$
are neutral Majorana fermions: heavy particles that arise naturally in
grand unified theories and may play a role in producing neutrino masses.
Both possibilities introduce new wrinkles because the fermion number
associated with the Majorana particles is no longer conserved and new
classes of diagrams must be considered.

If $\chi$ is the Majorana fermion, it does not change our estimates
significantly because it does not change the way that flavor flows through
the diagrams. The basic flavor backbone of the quark and $\Xi$ lines is
unchanged. There are simply more  ways  to connect the $\chi$ lines because
of their Majorana character. Our arguments about the flavor structure of CP
violating contributions are not affected, and the bounds on $\alpha_f$ and
$M_\chi$ are unchanged. 

If the $\Xi_x$ are Majorana fermions, our estimates are affected.
Generalized flavor is no longer conserved: there are no arrows on the
$\Xi$ lines. In particular, 
Higgs loops are no longer needed to obtain a non-zero
$\bar\theta$. We find contributions proportional to $K_{12}^2$, as defined
in (\ref{tb4}), from graphs like that shown in
figure~\ref{fig-8}.

{\figsize\begin{figure}[htb]
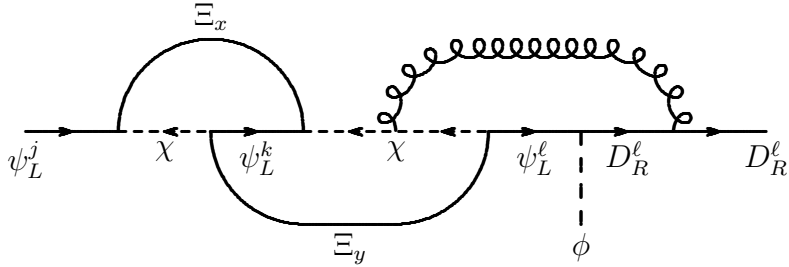

$$\beginpicture
\setcoordinatesystem units <\tdim,\tdim>
\stpltsmbl
\plot -150 -50 -200 -50 /
\setdashes <3pt>
\plot -150 -50 -100 -50 /
\setsolid
\plot -50 -50 -100 -50 /
\setdashes <3pt>
\plot -50 -50 50 -50 /
\setsolid
\plot 200 -50 50 -50 /
\tarrow from 170 -50 to 175 -50
\tarrow from 120 -50 to 125 -50
\tarrow from 70 -50 to 75 -50
\tarrow from 30 -50 to 25 -50
\tarrow from -20 -50 to -25 -50
\tarrow from -80 -50 to -75 -50
\tarrow from -120 -50 to -125 -50
\tarrow from -180 -50 to -175 -50
\setquadratic
\plot
150.0 -50.0
150.6 -46.2
152.2 -42.8
154.6 -40.6
157.1 -40.0
158.9 -41.0
159.4 -43.0
158.3 -44.8
155.9 -45.3
153.0 -44.2
150.3 -41.7
148.3 -38.4
147.4 -34.6
147.7 -30.8
149.1 -27.7
151.2 -26.0
153.4 -26.1
154.7 -27.7
154.5 -29.7
152.7 -31.2
149.7 -31.3
146.4 -30.1
143.4 -27.7
141.2 -24.5
140.0 -20.8
140.1 -17.3
141.4 -14.8
143.4 -13.9
145.4 -14.6
146.1 -16.6
145.2 -18.6
142.7 -19.9
139.2 -20.0
135.5 -18.8
132.3 -16.6
129.9 -13.6
128.7 -10.3
128.8  -7.3
130.3  -5.5
132.3  -5.3
133.9  -6.6
134.0  -8.8
132.3 -10.9
129.2 -12.3
125.4 -12.6
121.6 -11.7
118.3  -9.7
115.8  -7.0
114.7  -4.1
115.2  -1.7
117.0  -0.6
119.0  -1.1
120.0  -2.9
119.4  -5.4
117.2  -7.8
113.8  -9.4
110.0 -10.0
/
\plot
 40.0 -10.0
 36.2  -9.4
 32.8  -7.8
 30.6  -5.4
 30.0  -2.9
 31.0  -1.1
 33.0  -0.6
 34.8  -1.7
 35.3  -4.1
 34.2  -7.0
 31.7  -9.7
 28.4 -11.7
 24.6 -12.6
 20.8 -12.3
 17.7 -10.9
 16.0  -8.8
 16.1  -6.6
 17.7  -5.3
 19.7  -5.5
 21.2  -7.3
 21.3 -10.3
 20.1 -13.6
 17.7 -16.6
 14.5 -18.8
 10.8 -20.0
  7.3 -19.9
  4.8 -18.6
  3.9 -16.6
  4.6 -14.6
  6.6 -13.9
  8.6 -14.8
  9.9 -17.3
 10.0 -20.8
  8.8 -24.5
  6.6 -27.7
  3.6 -30.1
  0.3 -31.3
 -2.7 -31.2
 -4.5 -29.7
 -4.7 -27.7
 -3.4 -26.1
 -1.2 -26.0
  0.9 -27.7
  2.3 -30.8
  2.6 -34.6
  1.7 -38.4
 -0.3 -41.7
 -3.0 -44.2
 -5.9 -45.3
 -8.3 -44.8
 -9.4 -43.0
 -8.9 -41.0
 -7.1 -40.0
 -4.6 -40.6
 -2.2 -42.8
 -0.6 -46.2
 -0.0 -50.0
/
\springru 40 -10 *6 /
\setlinear
\circulararc 180 degrees from -50 -50 center at -100 -50
\circulararc 90 degrees from -100 -50 center at -50 -50
\plot -50 -100 0 -100 / 
\circulararc 90 degrees from 0 -100 center at 0 -50
\put {$\psi_L^j$} [t] at -200 -55 
\put {$\chi$} [t] at -125 -55
\put {$\psi_L^k$} [t] at -75 -55 
\put {$\chi$} [t] at 0 -55
\put {$\psi_L^\ell$} [t] at 75 -55 
\put {$D_R^\ell$} [t] at 125 -55 
\put {$D_R^\ell$} [t] at 200 -55 
\put {$\Xi_x$} [b] at -100 5 
\put {$\Xi_y$} [t] at -25 -105 
\put {~} at 0 0
\setdashes <5pt>
\plot 100 -100 100 -50 /
\put {$\phi$} [t] at 100 -105
\endpicture$$
\caption{\figsize\label{fig-8}A contribution to the Yukawa
coupling that produces a non-zero $\bar\theta$ with Majorana
$\Xi$s.}\end{figure}}

Graphs such as that in figure~\ref{fig-8} produce contributions to
$\bar\theta$ of order
\begin{equation}
{\alpha_s\over4\pi}\,\left({\alpha_f\over4\pi}\right)^2\,.
\label{mb1}
\end{equation}
This strengthens the bound on $\alpha_f$ by a factor of $4\pi v/m_t\approx18$.
For the special case of Majorana $\xi$s, the
the bounds on $\alpha_f$ and $M_\chi$ become:
\begin{equation}
\alpha_f<0.0024\,,\quad\quad M_\chi<1.2\times10^5\,{\rm GeV}\,.
\label{mb2}
\end{equation}
This model remains viable, although
$\alpha_f$ must be quite small to suppress $\bar\theta$. 
Nonetheless we find it 
quite attractive because the coupling of the light neutrinos to the
Majorana $\Xi$s could produce the neutrino masses seemingly
required to explain the recent evidence for oscillations
of atmospheric neutrinos~\cite{neutrinos}.\footnote{Equation (\ref{mb2})
yields too strong a bound for a plausible see-saw origin of neutrino masses.
However, the diagram in
figure~\ref{fig-higgs} involves two different $\Xi$s. If their masses are
hierarchical (with $m_\chi$ comparable to the lightest $\Xi$ mass),
(\ref{mb1}) is reduced by a $\Xi$ mass ratio and the upper bound on $m_\chi$
is correspondingly increased.} 

\section{Conclusions}

Wolfenstein's original ``superweak model''~\cite{wolfenstein} introduces CP
violation in an
{\it ad hoc\/} fashion.  We have reproduced the essential observable
consequences of this model with a minimal addition to the standard
model at a large mass scale. Other
potential solutions to the strong CP problem have been proposed, such as a
massless up quark or the Peccei-Quinn axion and its less visible variants.
These solutions are more elegant than ours, but are phenomenologically
challenged.  The observable consequence of our solution is the absence of
large CP violating phases in the Kobayashi-Maskawa matrix. Indeed, if such
a model is correct, direct CP violation in the $B$-sector will be difficult
if not impossible to detect and ongoing searches for a non-vanishing
$\epsilon^\prime$ parameter are bound to fail.

\bigskip
\centerline{\bf Note Added:\label{mele}}

In a recent preprint, S. Mele has argued that in a model like ours in which
the constraint from $\epsilon$
is removed, a real KM matrix can still fit the data~\cite{smele}. Checchia
{\it et.~al} \cite{checchia} come to the opposite conclusion, but they assume
that the new physics does not contribute significantly to the neutral $B$
meson mass difference (see \cite{hall}).

\bigskip\bigskip

\centerline{\bf Acknowledgments}\medskip

We are very grateful to D. Bowser-Chao and D. Chang for many important
comments.
One of us (HG) thanks Sheldon Stone for a helpful comment and gratefully
acknowledges the support of the Aspen Center for Physics, where some of this
work was completed.
One of us (SLG) thanks Paul Frampton for his challenge to devise a
plausible model of CP violation in which the KM matrix is nearly real. We are
also pleased to acknowledge conversations with S. Barr and R. Mohapatra.
{\em This work was supported in part by the National Science Foundation under
grant PHY-9218167.}

\appendix
\section{Right Handed Models}

Here we exhibit the three-loop diagram that contributes to $\bar\theta$ if the
CP violating heavy sector couples to the right-handed quarks rather than the
left-handed quarks. If, for example, the heavy sector couples to right-handed 
$D$ quarks, then the diagram in figure~\ref{fig-right} gives a nonzero
contribution to $\bar\theta$.
{\figsize\begin{figure}[htb]
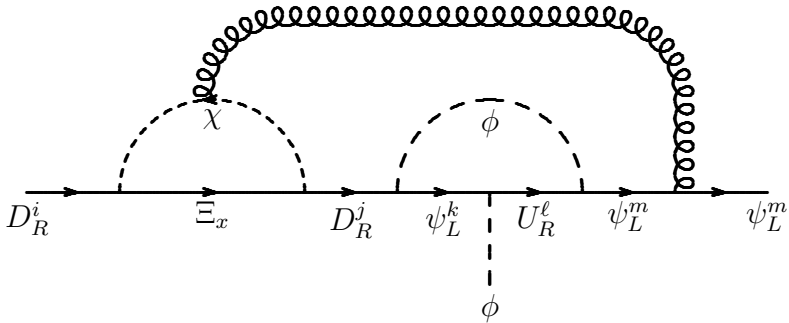

$$\beginpicture
\setcoordinatesystem units <\tdim,\tdim>
\stpltsmbl
\plot 0 -50 -400 -50 /
\tarrow from -28 -50 to -23 -50 
\tarrow from -78 -50 to -73 -50
\tarrow from -128 -50 to -123 -50
\tarrow from -178 -50 to -173 -50
\tarrow from -228 -50 to -223 -50
\tarrow from -303 -50 to -298 -50
\tarrow from -378 -50 to -373 -50
\tarrow from -298.5 0 to -304.5 0 
\setquadratic
\plot
-50.0   0.0
-48.7   3.9
-45.5   6.7
-42.0   7.0
-40.2   4.9
-41.5   2.5
-45.1   2.1
-48.7   4.2
-50.8   7.8
-50.3  11.9
-47.7  15.2
-44.3  16.2
-42.2  14.5
-43.0  11.9
-46.3  10.9
-50.4  12.2
-53.0  15.3
-53.4  19.4
-51.4  23.2
-48.3  24.8
-45.9  23.6
-46.2  20.9
-49.3  19.2
-53.5  19.7
-56.7  22.2
-57.9  26.2
-56.7  30.3
-53.9  32.5
-51.3  31.7
-51.1  29.0
-53.8  26.7
-58.0  26.4
-61.7  28.3
-63.6  32.0
-63.3  36.2
-61.0  38.9
-58.3  38.7
-57.5  36.1
-59.7  33.3
-63.8  32.1
-67.8  33.3
-70.3  36.5
-70.8  40.7
-69.1  43.8
-66.4  44.1
-65.2  41.7
-66.8  38.6
-70.6  36.6
-74.7  37.0
-77.8  39.6
-79.1  43.7
-78.1  47.0
-75.5  47.8
-73.8  45.7
-74.8  42.3
-78.1  39.7
-82.2  39.2
-85.8  41.3
-87.9  44.9
-87.5  48.5
-85.1  49.8
-83.0  48.0
-83.3  44.5
-86.1  41.3
-90.0  40.0
/
\plot
-260.0  40.0
-263.9  41.3
-266.7  44.5
-267.0  48.0
-264.9  49.8
-262.5  48.5
-262.1  44.9
-264.2  41.3
-267.8  39.2
-271.9  39.7
-275.2  42.3
-276.2  45.7
-274.5  47.8
-271.9  47.0
-270.9  43.7
-272.2  39.6
-275.3  37.0
-279.4  36.6
-283.2  38.6
-284.8  41.7
-283.6  44.1
-280.9  43.8
-279.2  40.7
-279.7  36.5
-282.2  33.3
-286.2  32.1
-290.3  33.3
-292.5  36.1
-291.7  38.7
-289.0  38.9
-286.7  36.2
-286.4  32.0
-288.3  28.3
-292.0  26.4
-296.2  26.7
-298.9  29.0
-298.7  31.7
-296.1  32.5
-293.3  30.3
-292.1  26.2
-293.3  22.2
-296.5  19.7
-300.7  19.2
-303.8  20.9
-304.1  23.6
-301.7  24.8
-298.6  23.2
-296.6  19.4
-297.0  15.3
-299.6  12.2
-303.7  10.9
-307.0  11.9
-307.8  14.5
-305.7  16.2
-302.3  15.2
-299.7  11.9
-299.2   7.8
-301.3   4.2
-304.9   2.1
-308.5   2.5
-309.8   4.9
-308.0   7.0
-304.5   6.7
-301.3   3.9
-300.0  -0.0
/
\springru -260 40 *16 /
\springdr -50 0 *4 /
\setlinear
\setdashes <3pt>
\circulararc -90 degrees from -350 -50 center at -300 -50
\circulararc 90 degrees from -250 -50 center at -300 -50
\setdashes <5pt>
\circulararc -90 degrees from -200 -50 center at -150 -50
\circulararc 90 degrees from -100 -50 center at -150 -50
\put {$D_R^i$} [t] at -400 -55 
\put {$D_R^j$} [t] at -225 -55 
\put {$\Xi_x$} [t] at -300 -55
\put {$\psi_L^m$} [t] at -75 -55 
\put {$\psi_L^m$} [t] at 0 -55
\put {$\phi$} [t] at -150 -105
\put {$\chi$} [t] at -300 -5
\put {$\phi$} [t] at -150 -5
\put {$\psi_L^k$} [t] at -175 -55 
\put {$U_R^\ell$} [t] at -125 -55 
\put {~} at 0 0
\setdashes <5pt>
\plot -150 -100 -150 -50 /
\linethickness=0pt
\putrule from 0 0 to 0 50 
\endpicture$$
\caption{\figsize\label{fig-right}A contribution to the Yukawa
coupling in models in which the CP violating heavy sector couples to the
right-handed $D$ quarks. }\end{figure}}
However, the corresponding diagram in our ``left-handed models'' that we
consider in this paper (figure~\ref{fig-left}), does not contribute to
$\bar\theta$. The crucial difference is that in our models, the only
potentially flavor-changing coupling of the right-handed quarks is the usual
coupling to the charged Goldstone boson components of the Higgs doublet, or
equivalently, to the longitudinal $W$s. As mentioned above, all diagrams
with an external $D_R$ must have a factor of $M_D$ associated with the $D_R$
external line. Thus the inverse mass matrices in (\ref{hc3}) cancel with this
explicit fact of the mass matrix, and the contribution of
figure~\ref{fig-left} to the determinant of the quark mass matrix is the trace
of a produce of
two hermitian matrices, and is therefore real. However, in
figure~\ref{fig-right},
the factor of $M_D$ occurs in the middle of the diagram, rather than on the
external line.

The three-loop contributions to $\bar\theta$ in right-handed models are not
huge. They may not be a phenomenological problem for some range of parameters.
But we hope that this discussion helps the reader see why there is no
contribution at all at the three loop level in our
left-handed models
(unless the $\Xi_x$ are Majorana particles, as discussed above).

{\figsize\begin{figure}[htb]
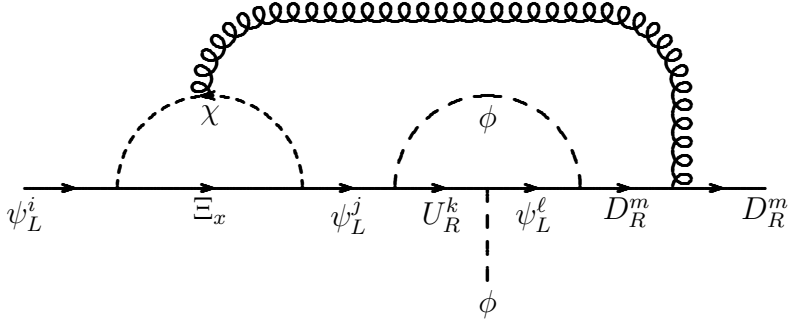

$$\beginpicture
\setcoordinatesystem units <\tdim,\tdim>
\stpltsmbl
\plot 0 -50 -400 -50 /
\tarrow from -28 -50 to -23 -50 
\tarrow from -78 -50 to -73 -50
\tarrow from -128 -50 to -123 -50
\tarrow from -178 -50 to -173 -50
\tarrow from -228 -50 to -223 -50
\tarrow from -303 -50 to -298 -50
\tarrow from -378 -50 to -373 -50
\tarrow from -298.5 0 to -304.5 0 
\setquadratic
\plot
-50.0   0.0
-48.7   3.9
-45.5   6.7
-42.0   7.0
-40.2   4.9
-41.5   2.5
-45.1   2.1
-48.7   4.2
-50.8   7.8
-50.3  11.9
-47.7  15.2
-44.3  16.2
-42.2  14.5
-43.0  11.9
-46.3  10.9
-50.4  12.2
-53.0  15.3
-53.4  19.4
-51.4  23.2
-48.3  24.8
-45.9  23.6
-46.2  20.9
-49.3  19.2
-53.5  19.7
-56.7  22.2
-57.9  26.2
-56.7  30.3
-53.9  32.5
-51.3  31.7
-51.1  29.0
-53.8  26.7
-58.0  26.4
-61.7  28.3
-63.6  32.0
-63.3  36.2
-61.0  38.9
-58.3  38.7
-57.5  36.1
-59.7  33.3
-63.8  32.1
-67.8  33.3
-70.3  36.5
-70.8  40.7
-69.1  43.8
-66.4  44.1
-65.2  41.7
-66.8  38.6
-70.6  36.6
-74.7  37.0
-77.8  39.6
-79.1  43.7
-78.1  47.0
-75.5  47.8
-73.8  45.7
-74.8  42.3
-78.1  39.7
-82.2  39.2
-85.8  41.3
-87.9  44.9
-87.5  48.5
-85.1  49.8
-83.0  48.0
-83.3  44.5
-86.1  41.3
-90.0  40.0
/
\plot
-260.0  40.0
-263.9  41.3
-266.7  44.5
-267.0  48.0
-264.9  49.8
-262.5  48.5
-262.1  44.9
-264.2  41.3
-267.8  39.2
-271.9  39.7
-275.2  42.3
-276.2  45.7
-274.5  47.8
-271.9  47.0
-270.9  43.7
-272.2  39.6
-275.3  37.0
-279.4  36.6
-283.2  38.6
-284.8  41.7
-283.6  44.1
-280.9  43.8
-279.2  40.7
-279.7  36.5
-282.2  33.3
-286.2  32.1
-290.3  33.3
-292.5  36.1
-291.7  38.7
-289.0  38.9
-286.7  36.2
-286.4  32.0
-288.3  28.3
-292.0  26.4
-296.2  26.7
-298.9  29.0
-298.7  31.7
-296.1  32.5
-293.3  30.3
-292.1  26.2
-293.3  22.2
-296.5  19.7
-300.7  19.2
-303.8  20.9
-304.1  23.6
-301.7  24.8
-298.6  23.2
-296.6  19.4
-297.0  15.3
-299.6  12.2
-303.7  10.9
-307.0  11.9
-307.8  14.5
-305.7  16.2
-302.3  15.2
-299.7  11.9
-299.2   7.8
-301.3   4.2
-304.9   2.1
-308.5   2.5
-309.8   4.9
-308.0   7.0
-304.5   6.7
-301.3   3.9
-300.0  -0.0
/
\springru -260 40 *16 /
\springdr -50 0 *4 /
\setlinear
\setdashes <3pt>
\circulararc -90 degrees from -350 -50 center at -300 -50
\circulararc 90 degrees from -250 -50 center at -300 -50
\setdashes <5pt>
\circulararc -90 degrees from -200 -50 center at -150 -50
\circulararc 90 degrees from -100 -50 center at -150 -50
\put {$\psi_L^i$} [t] at -400 -55 
\put {$\psi_L^j$} [t] at -225 -55 
\put {$\Xi_x$} [t] at -300 -55
\put {$D_R^m$} [t] at -75 -55 
\put {$D_R^m$} [t] at 0 -55
\put {$\phi$} [t] at -150 -105
\put {$\chi$} [t] at -300 -5
\put {$\phi$} [t] at -150 -5
\put {$U_R^k$} [t] at -175 -55 
\put {$\psi_L^\ell$} [t] at -125 -55 
\put {~} at 0 0
\setdashes <5pt>
\plot -150 -100 -150 -50 /
\linethickness=0pt
\putrule from 0 0 to 0 50 
\endpicture$$
\caption{\figsize\label{fig-left}The corresponding diagram in the
``left-handed models'' discussed in this paper does not contribute to
$\bar\theta$. }\end{figure}}


\begin{thebibliography}{99}
\bibitem{wolfenstein} L. Wolfenstein, Phys. Rev. Lett. {\bf 13} (1964)
562. For a recent review, see L. Wolfenstein, Comments Nucl.Part.Phys.21
(1994) 275. A different approach to the superweak interaction is discussed
in B. Holdom, 
Phys. Rev. {\bf D57} (1998) 357, {\tt hep-ph/9705231}.
\bibitem{jarlskog} C. Jarlskog, Phys. Rev. Lett. {\bf 55} (1985) 1039 and
Z. Phys. {\bf C29} (1985) 491.
\bibitem{strongcp} See: J.E. Kim, Phys. Rep. {\bf 150} (1987) 1 and
N. Ramsey, Phys. Scripta {\bf T59} (1995) 323.
\bibitem{bchao} D. Bowser-Chao, D. Chang and W.-Y. Keung, Phys. Rev. Lett.
{\bf 81} (1998) 2028. See also the useful review in D. Bowser-Chao, D. Chang
and W.-Y. Keung, {\tt hep-ph/9811258}.
\bibitem{hall} R. Barbieri, {\it et al.,} Phys. Lett. {\bf B425} (1998) 119,
{\tt hep-ph/9712252}.
\bibitem{neutrinos} T. Kajita, to appear in the proceedings
of the XVIIIth International Conference on Neutrino Physics and
Astrophysics, Takayama, Japan, June 1998; Super-Kamiokande collaboration,
{\tt hep-ex/9807003.}
\bibitem{smele} S. Mele, ``Indirect Measurement of the Vertex and Angles of
the Unitarity Triangle,'' CERN-EP/98-133, {\tt hep-ph/9810333}. 
\bibitem{checchia}
P. Checchia {\it et.~al}, ``A Real CKM Matrix?,'' DFPD-99-EP-4, Jan 1999. 7pp.
{\tt hep-ph/9901418}. 
\end{thebibliography}
\end{document}